\documentclass[lettersize,journal]{IEEEtran}
\usepackage{amsmath,amsfonts}
\usepackage{algorithmic}
\usepackage{algorithm}
\usepackage{array}
\usepackage[caption=false,font=normalsize,labelfont=sf,textfont=sf]{subfig}
\usepackage{textcomp}
\usepackage{stfloats}
\usepackage{url}
\usepackage{verbatim}
\usepackage{graphicx}
\usepackage{cite}
\usepackage[dvipsnames]{xcolor}
\usepackage{enumitem}

\newcommand{\kami}[1]{\textcolor{teal}{(Kami: #1)}}

\newcommand{\rev}[1]{\textcolor{olive}{#1}}

\begin{document}

\title{A Typology of Decision-Making Tasks \\ for Visualization}

\author{Camelia D. Brumar,
Sam Molnar,
Gabriel Appleby,
Kristi Potter, and
Remco Chang
\thanks{
Camelia D. Brumar and Remco Chang are with Tufts University, \\ e-mail: \textit{Camelia\_Daniela.Brumar@tufts.edu, Remco@cs.tufts.edu}

Sam Molnar, Gabriel Appleby and Kristi Potter are with the National Renewable Energy Laboratory, \\ e-mail: \textit{\{Sam.Molnar, GAppleby, Kristi.Potter\}@nrel.gov}
}
\thanks{Manuscript received September 9, 2024; revised April 5, 2025.}}

\markboth{Journal of \LaTeX\ Class Files,~Vol.~14, No.~8, August~2021}%
{Shell \MakeLowercase{\textit{et al.}}: A Sample Article Using IEEEtran.cls for IEEE Journals}

\IEEEpubid{0000--0000/00\$00.00~\copyright~2021 IEEE}

\maketitle

\begin{abstract}
Despite decision-making being a vital goal of data visualization, little work has been done to differentiate decision-making tasks within the field. While visualization task taxonomies and typologies exist, they often focus on more granular analytical tasks that are too low-level to describe large complex decisions, which can make it difficult to reason about and design decision-support tools. In this paper, we contribute a typology of decision-making tasks that were iteratively refined from a list of design goals distilled from a literature review. Our typology is concise and consists of only three tasks: CHOOSE, ACTIVATE, and CREATE. Although decision types originating in other disciplines exist, we provide definitions for these tasks that are suitable for the visualization community. Our proposed typology offers two benefits. First, the ability to compose and hierarchically organize the tasks enables flexible and clear descriptions of decisions with varying levels of complexities. Second, the typology encourages productive discourse between visualization designers and domain experts by abstracting the intricacies of data, thereby promoting clarity and rigorous analysis of decision-making processes. We demonstrate the benefits of our typology through four case studies, and present an evaluation of the typology from semi-structured interviews with experienced members of the visualization community who have contributed to developing or publishing decision support systems for domain experts. Our interviewees used our typology to delineate the decision-making processes supported by their systems, demonstrating its descriptive capacity and effectiveness. Finally, we present preliminary findings on the usefulness of our typology for visualization design. 
\end{abstract}


\section{Introduction}
Visualization researchers and practitioners alike have suggested that one of the primary goals of visualization systems is to help users make decisions\cite{sacha2014knowledge, thomas2006visual, ware2019information, dimara_critical_2022}.
Many works have come out of the visualization community that have succeeded in helping users make important decisions (e.g., \cite{afzal2011decisionsupportforepidemic, waser2014manyplans, chang2007wirevis, Alameldin94,ferreira2015urbane,gaba2023my}, to name a few). 
However, it is difficult to understand patterns across these systems, even among those that address similar decision-making problems in the same domain.
Furthermore, the taxonomies within the visualization community (e.g., \cite{amar2005low, yi2007toward, brehmer_multi-level_2013, dimara_critical_2022, dimara2021unmet, oral_information_2023}) do not adequately capture the components of complex decision-making problems and processes because they instead focus on analytical tasks or visual encodings, or they categorize all decision-making tasks under a single ``decide'' or ``choice'' task.
This leaves a critical gap in understanding how information is processed when making a decision and how individual decisions combine to form a coherent decision-making process.

\begin{figure}[t!]
\centering  \includegraphics[width=\linewidth]{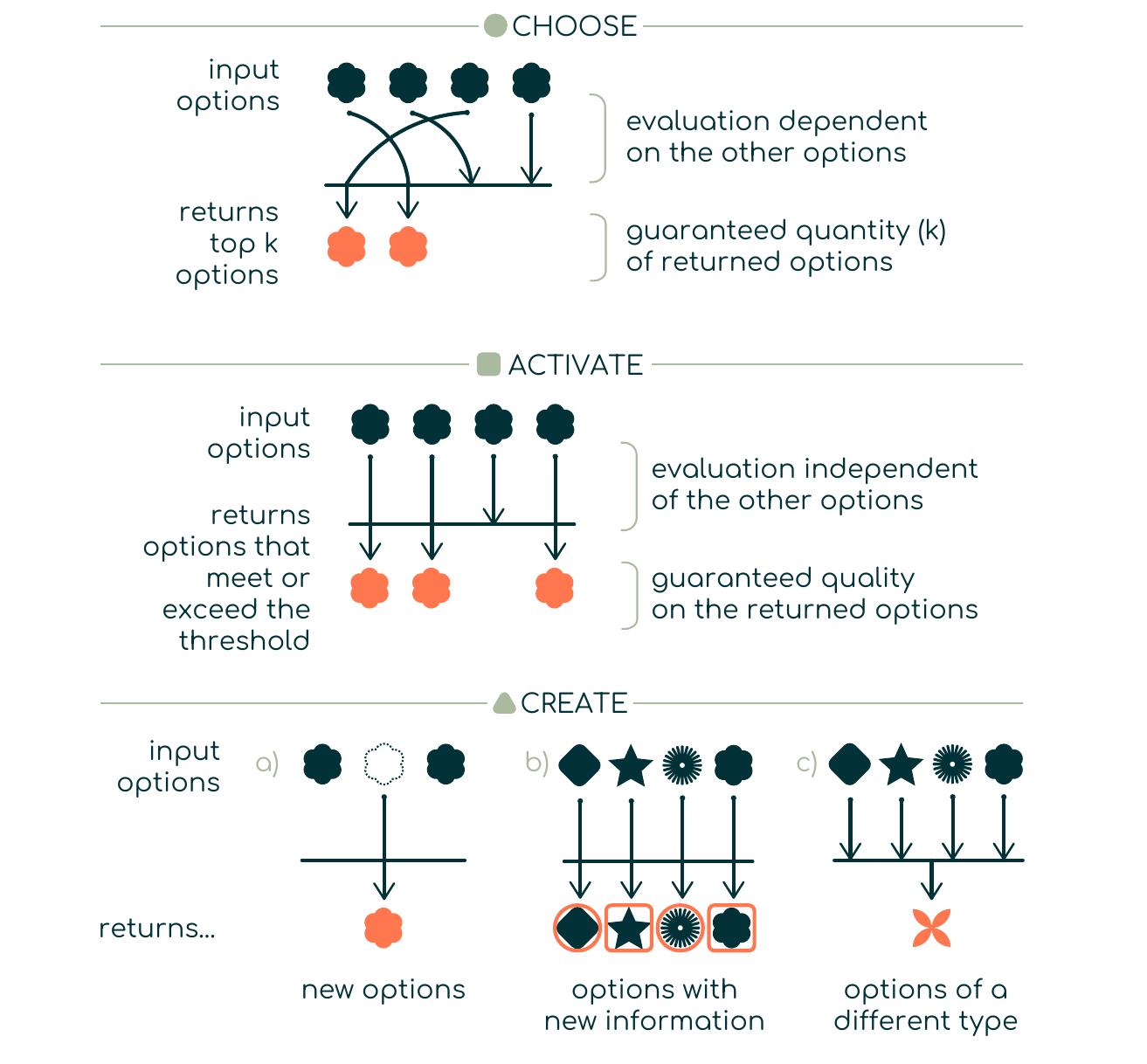}
  \caption{The three decision tasks: CHOOSE, ACTIVATE, and CREATE.}
  \label{fig:teaser}
\end{figure}

To address these shortcomings, we introduce a typology specifically focused on decision-making tasks in visualization.
It consists of three decision tasks: CHOOSE,  ACTIVATE, and CREATE, as shown in Figure~\ref{fig:teaser}. 
Each task assesses options and represents distinct decisions with those options.
The CHOOSE task decides which options are optimal or best, the ACTIVATE task decides which options meet or exceed a threshold, and the CREATE task decides how to assemble, synthesize, or generate new information from the options.
The most critical properties of this typology are that it allows us to represent the complex structures of real-world decision-making problems through the hierarchical organization of tasks and the composition of tasks.

We designed our typology using design goals derived from a literature review of decision-support tools within the visualization community.
We built upon previously curated corpora of application and decision-support visualization systems\cite{oral_information_2023, shi_how_2023}, focusing on those with domain expert evaluations.
We then followed an inductive coding process to identify each paper's primary decision-making goals and processes.
From these codes, we distilled the design goals for our decision-making typology, including iterativeness, composability, generalizability, simplicity, and others.
From decision-making theory\cite{fischhoff_communicating_2014, simon1960new} and our design goals, we worked towards an initial set of tasks for the typology.
We used our survey to hone and iterate on our tasks until they could accurately characterize the works included.
To illustrate that our typology has the aforementioned properties, we present four case studies where authors described the decision-making processes addressed by their tools using our typology.

We validated our typology by conducting semi-structured interviews. 
Our nine participants were visualization experts from academia, government laboratories, and industry who had authored one or more decision-support tools.
We demonstrated our typology to the participants and asked them to apply it to their systems and answer open-ended questions about the typology, its properties, and how it compares to other typologies.

We divide the results of our validation into three broad categories: the correctness of the typology, the utility of the typology, and how the typology compares to existing task taxonomies and typologies.
The correctness of the typology can be further broken down into the consideration of completeness (the typology needs to cover all decision-making problems we have encountered) and expressiveness (the typology can be easily used to describe decision-making scenarios).
In our interview study, participants found that the typology is complete regarding the three decision tasks and that they could successfully express their decision-making problems in diagrams using the typology.
They also found the typology helpful when analyzing and reflecting on the design process of visualization decision-support tools, believed that it could be useful when communicating with domain experts, and noted that it could categorize visualization decision-making publications.
Moreover, participants provided valuable insights when comparing the typology to existing frameworks, highlighting its appropriate level of abstraction for describing decision-making problems and its adaptability to diverse data and contexts.

Lastly, we conducted a preliminary study to evaluate the impact of our typology on visualization design. Our findings revealed that participants used the typology to guide their choices between design alternatives, identify improvements for decision tasks that lacked sufficient support, and even redesign their systems based on the insights gained from applying the typology. With the results from our evaluation, we gathered all the insights as limitations and possible future avenues for our typology.

In summary, the major contributions of this paper are:
\begin{itemize}
    \item Design goals for decision-making typologies derived from the thematic analysis of a comprehensive literature review.
    \item A typology of decision tasks, which provides an abstract yet precise way to discuss decision-making activities in visualization.
    \item Four case studies in which our typology is used to describe users' decisions in the context of decision-support tools that demonstrate the properties of our typology.
    \item A semi-structured interview study that serves as an evaluation of our typology and provides a roadmap for future work.
    \item A preliminary study on the design implications of our typology and discussion on the utility of this typology in the visualization community.
\end{itemize}
\section{Background \& Related Work}
Decision making is a broad topic studied across many disciplines, and visualization researchers emphasize that supporting decision making is a vital goal of data visualization~\cite{dimara_critical_2022, hullman_decision_2024, amar2005knowledge, card1999readings, andrienko2006exploratory}.
To develop effective support, the visualization community utilizes typologies as tools to reason about visualization designs and explain problems and user tasks addressed by visualization. 
We review the relevant literature from these areas and discuss the significance to this work.

\subsection{Decision Making \& Visualization}
The visualization community tends to focus on decision making from three broad lenses: cognition and perception, decision process support, and domain-specific decision support systems.
In a cognitive context, decision-making is often framed in terms of rationality and judgement\cite{padilla_decision_2018}.
In this view, the most rational decisions are ones that maximize expectation of a function\cite{fischhoff2020judgment}.
However, a number of human decisions are not rational\cite{takemura2021behavioral}, a fact that has impacted the research proclivities of the visualization community in two significant ways.
The first is an increase in research on computational decision-making support~\cite{wang2021survey,hullman_decision_2024,wu2021ai4vis}.
While promising, it is unclear and often difficult to determine under what conditions users will accept help from or implement decisions of computational models~\cite{padilla2022multiple,gaba2023my}.
The second impact is the increased study of cognitive bias and perception in visualization contexts.
Cognitive biases explain irrational decision-making as the result of using heuristic functions to evaluate information in a decision-making cognitive process\cite{dimara2018task}.

The occurrence of specific kinds of judgment and decision-making errors can then be traced to specific heuristic functions, and recent studies have investigated the overlap of specific biases with visualization design\cite{dimara2016attraction}.
There is evidence that visualization design can impact the strength of a cognitive bias\cite{dimara2016testing}, and there exists a significant body of work that assesses the role of visual perception in judgment and decision-making\cite{hosseinpour2024examining,padilla_decision_2018,oral2023decoupling}.
For example, in Padilla et al.\cite{padilla2022impact}, the authors find a significant relationship between participants' risk assessment and uncertainty visualization design choices.
While important, cognitive and perceptual approaches focus on reasoning and judgement, rather than describing decision-making problems in domain agnostic contexts for meaningful discussions on visual support. 

A second body of work in the visualization community is concerned with designing visualizations that support specific problem-solving structures for a decision-making problem.
These visualization systems are unique in that they support a specific approach for decision-making rather than focusing on a specific application or domain\cite{pajer2016weightlifter,gratzl2013lineup,wall2017podium,asahi1995using,oral_information_2023}.
One such decision process addressed in the community is multiple criteria decision making (MCDM) which is a mathematical formalism that enables selection of a solution given multiple competing criteria and a decision maker's preferences for various attributes of a solution\cite{bozorg2021handbook}.
MCDM lends itself nicely to computation and is particularly useful for ranking solutions\cite{henig1996solving}, defining (and therefore exploring) trade-offs among solutions\cite{pajer2016weightlifter}, and the explicit use of a person's preferences can be an invaluable source of information for recommendation systems or group decision-making\cite{hindalong2020towards}.
However, a prescribed structure to a decision-making problem is not always appropriate, and transforming a decision-making problem to fit into a particular structure can itself be challenging for users.
For example, MCDM hinges on the specification of criteria and evaluation functions for those criteria\cite{bozorg2021handbook}.
However, for some problems the selection of criteria and evaluation function is in itself difficult\cite{pajer2016weightlifter,khannoussi2022simple,olteanu2022preference}, and externalizing preferences as abstract values can be a hard interpretation problem for users\cite{kuhlman2019evaluating}.
Focusing solely on describing and discussing problems from a specific structure unnecessarily constrains possible visualization solutions, and does not capture the diverse ways people can approach complex decision-making problems.

Finally, there exists a large body of work in the visual analytics community which addresses decision-making from domain-specific contexts.
For example, many decision support systems exist to address medical decisions\cite{kreiser2017decision,van2014comparative,Alameldin94}, security and fraud detection\cite{wang2020umbra,chang2007wirevis,Nguyen2019,scheepens2015rationale}, city planning\cite{aseniero2015stratos,zhao2017skylens,ferreira2015urbane}, and other domain-specific applications\cite{rudolph2009finvis,azuma1999,sorger2015litevis,waser2014many,booshehrian2012vismon}.
When supporting decision making within a particular domain, it is indeed important to take into account the nuances relevant to that domain.
However, domain-specific approaches do not necessarily generalize, and do not allow for a discussion of decision-making problems across domains.

\subsection{Visualization Typologies}

Decision-making happens at many different levels of interaction with data and within visualization systems.
Yet, no existing visualization typology can capture both high-level decisions (e.g., what car to buy) and low-level decisions (e.g., where do I click next in the application to show cars with specific attributes?) while simultaneously representing relevant decisions that occur outside of those systems. 
We briefly review prominent and relevant typologies of the visualization community here and discuss why these typologies are insufficient for discussing decision-making.

Two of the most prominent and influential visualization task typologies in the community are the analytical task typology by Amar et al.~\cite{amar2005low} and the multi-level task typologies by Brehmer and Munzner~\cite{munzner2014visualization,brehmer_multi-level_2013}.
The analytical tasks typology addressed a gap in highly data-centric typologies by taking an analytical-centric approach.
This typology consists of ten tasks that describe ``primitive" \textit{analyses of data}, or low-level tasks~\cite{amar2005low}.
The multi-level task typology further expanded upon these tasks to incorporate higher levels like ``why?", ``how?", and ``what?", which can be composed from lower-level tasks~\cite{munzner2014visualization,brehmer_multi-level_2013}.
However, these tasks are not suitable for describing decision-making in compact yet meaningful ways. 
Consider the decision of buying a car by choosing among a set of available used cars.
This task, in its current form, cannot yet be represented in either the analytical or multi-level task typologies because more information is needed on the analytical activities and data.
In other words, these typologies are oriented towards describing potential \textit{solutions} or approaches for solving a problem, they do not describe the decision \textit{problem} or how decisions are made.
When we can describe decision problems in abstract yet precise ways, we can begin to reason about solutions more holistically.

\section{Methodology}
\label{sec:methodology}

This section describes our methodology for developing a typology of decision-making tasks. 
Our approach involved gathering design goals from a literature review of publications in the following fields: visualization relating to decision-support tools, cognitive psychology relating to mental processes for decision-making, and other scientific literature relating to decision-making.
Specifically, we iteratively analyzed design studies and visual analytics application papers to extract example scenarios and relevant contexts relating to decision-making, as detailed in subsections \ref{subsec:corpus} and \ref{subsec:coding}.
This analysis led to a set of design goals we believe our typology should fulfill in order to accurately represent the decision processes described in the corpus’s case studies and decision-making scenarios.
These design goals are described in section \ref{subsec:desiderata}.
Next, we shaped the typology using theory frameworks for decision-making by keeping in mind the goals we derived from the application papers, as explained in section \ref{subsec:design}.
The result of this design process is our proposed typology of decision-making tasks for visualization, which we present in detail in Section \ref{sec:typology}.

\subsection{Application Survey Corpus} \label{subsec:corpus}
We reviewed relevant visualization decision-support literature to formulate the design goals of our typology, building upon the literature corpora curated by Oral et al.\cite{oral_information_2023} and Shi et al.\cite{shi_how_2023}. Oral et al.'s corpus focused on papers explicitly labeled as decision-making tools, filtering based on the presence of the term `decision' in titles or abstracts. However, this approach omitted valuable visualization papers which serve as decision-support tools for domain experts and tools implementing methodologies other than MCDM. To broaden the scope, we incorporated Shi et al.'s work, allowing us to encompass a more comprehensive range of visual analytics applications and design studies. From their list of 190 VAST papers\cite{shi_how_2023}, we found 63 to be applications with a direct decision-making focus.
We added to that list the papers with ``decision'' mentioned in titles or abstracts to be consistent with\cite{oral_information_2023}, adding up to 74 papers.

To refine our selection, we prioritized papers featuring specific domain applications, and eliminated generic systems lacking evaluations with domain experts. This guaranteed that the chosen tools were utilized by domain users with actual decisions and requirements that could be validated. For instance, papers solely based on hypothetical case studies were excluded (e.g., Podium\cite{wall2017podium}, Zooids\cite{le2016zooids}).
This filtering process resulted in a final corpus of 69 papers that directly contribute to our study's objectives, and is available in the Supplemental Materials. 

\subsection{Coding Process} \label{subsec:coding}
Our coding process followed an inductive approach to derive the design goals for our typology from the survey corpus outlined in the preceding section. 
The process involved a team-based approach with two coders, who are the first two authors of this paper. 
For each paper, they determined the overall decision goal by analyzing the task abstraction, requirements, and evaluation sections, as well as any supplementary materials, such as figures or video demonstrations (if available).
The identified decision goals of the tools informed the design goals for a task typology.
In addition, we noted unique decision contexts, such as sub-decisions that supported the overall decision, and the information that decisions use and produce.
The result of this coding process resulted in seven design goals for our typology, which are discussed in the next subsection.

\subsection{Design Goals for a Decision Task Typology} \label{subsec:desiderata}
Through multiple brainstorming sessions involving all authors, we distilled the following list of goals that we believe a decision-making typology for visualization should fulfill in order to accurately represent the decision processes and scenarios found in the corpus.
In the supplemental materials, we include excerpts from the corpus of papers that support the choice of each goal.

\begin{enumerate}[label=\textbf{G\arabic*}]
    \item \label{goal:iteration} \textbf{Iteration}: Given that decision-making processes in visualization often involve cycles and iterations, our typology should incorporate a mechanism to express such iterative patterns.
    
    \item \label{goal:hierarchical} \textbf{Hierarchical Structure}: Decisions are intricate and can be deconstructed into subtasks, necessitating the ability of our typology to capture these hierarchical relationships.
    
    \item \label{goal:process-centric} \textbf{Process-Centric Approach}: Our typology should focus on delineating the sequence of decisions users encounter, rather than merely representing data-processing flow or interaction order typically outlined in task lists and design requirements of application papers and design studies.
    
    \item \label{goal:data-flexibility} \textbf{Information Processing-Focused}: 
    The decision tasks within our typology should demonstrate flexibility and impartiality towards input and output types, instead emphasizing how information is processed and evaluated.

\end{enumerate}

Additionally, we define the following evaluation criteria, which we use to assess the effectiveness of our typology through user evaluations:
\begin{enumerate}[label=\textbf{E\arabic*}]
    \setcounter{enumi}{0}
    \item \label{criteria:generalizability} \textbf{Generalizability}: The typology should apply uniformly across the diverse range of papers we surveyed and should remain relevant across various domains.
    \item \label{criteria:simplicity} \textbf{Simplicity}: We aspire for our typology to remain straightforward and concise, with minimal decision types and operations to avoid unnecessary complexity.
    \item \label{criteria:accessibility} \textbf{Accessibility}: Our typology should be easily understood and grasped by both the visualization designers and the domain experts, contributing to its usability and adoption.
\end{enumerate}

These criteria guided the creation of our typology, as we discuss in the following subsection.

\subsection{Designing the Typology}\label{subsec:design}

Our design process began with an initial set of decision-making tasks inspired by relevant decision-making theory literature, particularly works by Fischhoff et al.\cite{fischhoff_communicating_2014,fischhoff2020judgment} and Simon's decision-making phases \cite{simon1960new,campitelli2010herbert}.
Then, using the goals from the previous subsection and other theory found in cognitive science literature\cite{wang2007cognitive,padilla_decision_2018,takemura2021behavioral,dimara2018task}, we iteratively added, removed and modified the decision tasks and their properties until we achieved a stable set of decision-making tasks and definitions, resulting in the finalized typology described in Section \ref{sec:typology}.





In terms of the different types of decision-making tasks, previous works often describe a single decision task, \textit{decide} \cite{dimara2018task, dimara_critical_2022} or \textit{choice} \cite{wang2007cognitive, dimara2017conceptual, oral_information_2023}, both referring to the action or process of deciding or resolving something.
While useful for encapsulating decisions as a whole, they lack descriptiveness regarding the type of decision, that is, \textit{how} a decision is made, limiting the analysis of decision-making in visualization\cite{dimara2018task}.

There are two prominent works that are relevant to consider which outline possible decision tasks.
We tested each of the decision tasks outlined in these works by applying them to our corpus.
For example, we conducted an initial pass over the decision-making support tools and labeled them with one of the decision tasks, attempting to represent the ultimate decision the tool supports.
For each categorization, we noted which of our design goals they did and did not meet.

First, Fischhoff et al.\cite{fischhoff_communicating_2014} describe three distinct decisions in the form of questions: ``Is it time to act?", ``Which is best?", and ``What is possible?". While these better describe types of decisions in a general and domain-agnostic way (\textbf{\ref{criteria:generalizability}}), the associated definitions are not directly applicable to visualization because they lack specificity on the use of information to reach conclusions or achieve decision goals (\textbf{\ref{goal:data-flexibility}}), which we found distinguished decisions and visualization designs in our corpus.
Additionally, as defined by Fischhoff et al., these tasks are not composable (\textbf{\ref{goal:iteration} - \ref{goal:process-centric}}), which limits our ability to describe complex decision-making processes in sufficient detail for designing decision-support systems (\ref{criteria:generalizability} - \ref{criteria:accessibility}).

The second prominent work outlining decision types is by Herbert Simon, which describes decision-making as a process comprising three stages: \textit{intelligence}, \textit{design}, and \textit{choice}.
The \textit{intelligence} stage involves gathering information and identifying problems or opportunities that require a decision, the \textit{design} stage involves developing, analyzing, and evaluating options, and the \textit{choice} stage involves selecting and implementing one of the options, considering constraints and objectives.
According to Simon, each of these stages can be further broken down into sub-stages of the same type.
While Simon's model meets goals \textbf{\ref{goal:iteration} - \ref{goal:hierarchical}} and has property \textbf{\ref{criteria:generalizability}}, it does not specify how information is processed within these stages (\textbf{\ref{goal:data-flexibility}}) or how the stages interrelate to achieve the final decision (\ref{goal:process-centric}).


Fischhoff and Simon's decision types meet some of our typology design goals, but both fail to delineate differences in information processing for decisions, and neither discuss the composition of decisions in sufficient detail for visualization.
Knowing the specifics of how information is processed, and therefore how decisions are made, and how smaller decisions compose together to reach a final decision, allows designers to create visualizations that better support those processes.
We expand on past work by explicitly addressing how information is processed in our tasks (\textbf{G1-4}), and ensure the aforementioned properties exist in our typology, which we discuss next.

\section{Typology of Decision-Making Tasks} \label{sec:typology}
Our typology consists of three tasks, which we derived from examining the scientific literature and iteratively refining the definitions to meet our design goals (see Section \ref{sec:methodology}).
These tasks reflect and capture the ubiquity of decision-making, flexibly describe the execution of decisions, and the definitions use language consistent with the decision-making literature.
The tasks are agnostic to who or what performs the evaluation, which ensures the typology can describe human or computational decision-making problems.
Furthermore, our task definitions are \textit{agnostic to the rationality} of the evaluations, which allows for our typology to be consistent and used in conjunction with computational and cognitive biases\cite{dimara2018task, mehrabi2021survey}.
We illustrate each decision task visually in Figure~\ref{fig:teaser}.

\subsection{Task Definitions}
Each task in our typology evaluates and returns options according to its definition.
\textit{Options} are information entities typically relevant for a decision task.
Options can have \textit{features} that describe characteristics of the options.
Lastly, \textit{criteria} are preferences or standards on features of the options. 
As an example, we could consider options as cars, with features of color, price, and mileage, and 
an example criteria is ``any color except white".

\subsection*{\includegraphics[height=8pt]{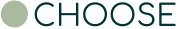}}


The CHOOSE task involves assessing a set of options and returning a subset deemed optimal or best.
Fischhoff’s concept of \textit{decisions with fixed options}, and Simon’s final decision-making stage, \textit{choice}, form the basis of the CHOOSE task.
We more specifically define CHOOSE as deciding on the top or best $k$ options available (illustrated at the top of Figure~\ref{fig:teaser}).
Importantly, within CHOOSE, the evaluation of options is \textit{dependent} on the other options.
Whether an option is returned depends on how it compares with the other available options.

Consider the example decision of buying a car; we have four cars from which we can CHOOSE.
Rather than looking for a car that satisfies a set of criteria, we are looking for the best car \textit{from the current set available}.
Furthermore, the notion of \textit{best} can change over time, which can lead to different choices.
These changes could be due to different environmental or circumstantial contexts (\textit{e.g.}, living in a new location), or shifting perceptions (\textit{e.g.}, brand loyalty).
Additionally, cognitive biases like loss aversion or the framing effect~\cite{dimara2018task}, and machine biases like the linking bias~\cite{mehrabi2021survey}, are examples of biased evaluations that can occur in the CHOOSE task.

Unlike the other tasks in our typology, with CHOOSE we know how many options are returned upon execution, meaning only one execution of the CHOOSE task is required to reach a subset of options of size $k$.
Although criteria can inform comparison of options in CHOOSE, criteria are not necessary to execute CHOOSE.
Furthermore, it is not a given that the returned options meet a set of criteria.
In other words, we have \textit{no guarantees on the quality} of the returned options, only that they are the best subset of all evaluated options.

\subsection*{\includegraphics[height=8pt]{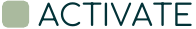}}


The ACTIVATE task draws inspiration from Fischhoff’s decisions about action thresholds but expands beyond merely temporal thresholds.
This task represents a decision where options are evaluated, and only those that meet or exceed a threshold are returned (illustrated in the center of Figure~\ref{fig:teaser}).
In contrast to CHOOSE, the evaluation of one option is \textit{independent} of the others.
Whether an option is returned depends only on its evaluation individually, regardless of the evaluation of the other options.

In the car-buying example, the ACTIVATE task can be represented by applying thresholds to each car such that only cars under a particular price and mileage are left after evaluation.
Notice that because each car is evaluated independently, we can assess them all in the evaluation \textit{simultaneously}.
There is flexibility in the evaluation that occurs in ACTIVATE, as thresholds need not be strictly enforced across options or well-defined.
For example, in the car buying decision, a person may exclude a car that is very near (but still under) \textit{both} the price and mileage limits, but keep a car that is slightly over the price limit because the mileage is much lower than the limit.
Some example biases that can occur in this evaluation include the cognitive biases of information bias and the anchoring effect\cite{dimara2018task}, or the algorithmic processing bias in computational decision-making\cite{danks2017algorithmic}.

Unlike CHOOSE, where we know the exact number of options that are returned, we do not have this guarantee for ACTIVATE.
It is possible that none or all options are returned by ACTIVATE.
We have to execute ACTIVATE an unknown number of times to reach a subset of options of size $k$.
However, any options returned by ACTIVATE meet a level of acceptability that is informed by given criteria.
In other words, ACTIVATE can \textit{guarantee a level of quality}  on the returned options that is not possible to do with CHOOSE.

\subsection*{\includegraphics[height=8pt]
{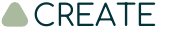}}
\label{sec:typology-create}



The CREATE task represents decisions on assembling, synthesizing, or generating new information.
This decision represents the creation of knowledge or insights\cite{battle2023we, chang2009defining, keim2008visual, sacha2014knowledge}, or more tangibly, reports\cite{chinchor2009science}, new data transformations\cite{kasik2009data}, etc.
In contrast with the other tasks in our typology, CREATE can take any number of options (including none) and return new options (\textit{e.g.}, adding to the set of options or new features to existing options), options with new information or modifications, or returning a set of options of a completely different type. 
These facets of the CREATE task are illustrated as a, b, and c, respectively, at the bottom of Figure \ref{fig:teaser}.

Consider again the decision to buy a car.
Rather than deciding on an existing set of cars, the CREATE decision allows for three possibilities: 
(1) the addition of a new car that is not in the current set of options, 
(2) the modification or inclusion of features such as adding whether each car has advanced driver assist technologies, or 
(3) synthesizing new information based on the options, like creating a custom car where the initial set of options are components of a car (\textit{e.g.}, body, wheels, software capabilities).
The last possibility is perhaps the most impactful, as it describes the case where the evaluation in CREATE returns a new type of option than the initial set of options. 

To further illustrate (3), consider an evacuation planning scenario faced by emergency responders during a wildfire. One task involves using predicted fire spread (the input information) to determine which households should receive evacuation orders. While this could be framed as a task involving refining a list of all households (i.e., a CHOOSE or an ACTIVATE task), the decision can also be reframed as a CREATE task: evaluating one type of information (e.g., spatiotemporal fire forecasts) to generate a new type of actionable output (e.g., a list of households to evacuate).
This example of transformation, from predictive modeling to actionable directives, highlights another core capability of the CREATE task: the output is not necessarily a refined version of the input but may be a fundamentally different representation, tailored to a different decision context. As discussed further in Section \ref{subsec:discussion-of-definitions}, the ability to reframe a decision as different decision-making tasks is a strength of the typology, as it supports reflection on the various ways information can be processed.



The CREATE task is often considered a precursor to decisions in the literature. 
In Simon’s framework, creating alternative courses of action by evaluating their consequences falls under the \textit{design} stage \cite{simon1957background, simon1960new}. 
However, in agreement with Fischhoff et al. \cite{fischhoff_communicating_2014}, we argue that the CREATE task is more than just a precursor to a decision-- it is itself a fundamental decision that can be the ultimate goal of a decision support tool. 
For instance, generating a report that synthesizes insights while using a tool exemplifies how the CREATE task serves as an end decision goal.

While we can make some guarantees on the quantity or quality of returned options in CHOOSE and ACTIVATE, we cannot do so for CREATE.
If the initial options do not meet a set of criteria, this can \textit{propagate} through the CREATE evaluation and impact the \textit{quality} of the returned options.
Furthermore, even if the initial options do meet a set of criteria, the returned options may require a different set of criteria in order to assess their quality.
Constraints on the \textit{quantity} of returned options from the CREATE evaluation could arise from initial options that do not meet a set of criteria.
However, the occurrence of inspiration, innovation, or randomness in the CREATE evaluation can allow an unbounded number of returned options.
The quality and quantity assertions for CHOOSE and ACTIVATE are possible because we can reason directly on the initial options in relation to their evaluations.
While we can also reason about the impact of the initial options on the CREATE task, the CREATE evaluation is less bounded in that there is no prescribed way to use the initial options.
This leads to additional uncertainty in the quality and quantity of returned options than what is present in CHOOSE and ACTIVATE.



\subsection{Discussion of Definitions}
\label{subsec:discussion-of-definitions}

The three decision tasks are distinguished by how they evaluate information, or \textit{how} a decision is made, thus meeting goal \textbf{G4}.
The same decision \textit{goal} (\textit{e.g.}, buying a car) is achievable with any decision \textit{task} if the appropriate information (\textit{e.g.}, the cars, their features, and the criteria) is available and the evaluation is possible.
For example, a user could prefer an ACTIVATE task because evaluating options individually by their features is easier than assessing options against one another.
Alternatively, another user may prefer to execute a CHOOSE task because assessing an option in the context of the available options is helpful.
Or perhaps yet another user prefers to execute a CREATE task because they prefer to devise their own option while drawing inspiration from the available ones.
Although we use clear language distinguishing options from features and criteria, the decision tasks can operate on any of these entities.
For example, as part of the decision goal of buying a car, we could CREATE the criteria that are then used to evaluate cars.
The decision tasks allow us to describe how different users or algorithms approach decision-making, as well as how their strategies may shift in response to evaluation challenges or changes in information availability.

The decision tasks in our typology are independent from but synergistic with the intent (task framing), judgement (valuation of information), or analytical actions that can occur while executing the task evaluation.
Specifically, these facets of decision-making do not change the \textit{evaluation mechanics} of the decision tasks, but they can impact the \textit{outcome} of a decision task.
For example, an ACTIVATE task does not depend on whether a user executed it with the intent to include the best options or to exclude the worst options, as either intent can be achieved with an ACTIVATE evaluation.
However, the intent may influence how each option is assessed and therefore the outcome of the ACTIVATE task.
This distinction is further highlighted when considering the judgment and analysis of information.
An anchoring bias in a CHOOSE task influences how later options are evaluated, but it does not change the details of the evaluation.
Whether we use sorting or correlation as an analytical technique in CHOOSE only impacts which options are returned.
Our decision tasks describe the unique functions of how decisions are made, while these other facets of decision-making help explain potential differences in decision outcomes.

The distinction on how information is processed to make a decision, rather than what influences the outcome of a decision, helps ensure our typology is simple, generalizable, and accessible (properties \textbf{\ref{criteria:generalizability} - \ref{criteria:accessibility}}).
Given our typology has the flexibility to describe a decision goal using any decision task, or combination of decisions tasks, hints at other important properties of our typology that help us meet goals \textbf{\ref{goal:iteration} - \ref{goal:process-centric}}, namely the composability and hierarchical structure of the tasks.

\section{Using the Typology for Decision-Making Problems}
\label{sec:composability}
In isolation, each decision task already provides new ways to express decisions in relevant ways for visualization.
However, many decision problems are complex and involve multiple related decisions\cite{simon1957background}.
While a list of sub-decisions can partially describe larger decisions, it cannot capture the structural relationship that clearly exists among decisions.
In order to meet goals \textbf{\ref{goal:iteration} - \ref{goal:process-centric}}, we determined how our decision tasks can represent more complex and interrelated decisions.
Specifically, the decision tasks are \textit{functional descriptions} of information evaluation, which exposes two properties of our typology that enable us to combine tasks and represent decision-making problems: composability and divisibility.
Furthermore, we can use a diagram to illustrate how these properties can describe more complex decision problems.

\subsection{Properties that help describe decision problems}

Consider again the example decision of wildfire evacuation: CHOOSE an evacuation location for residents.
We could decompose this CHOOSE task into several sub-decision tasks like a CREATE task that decides a possible set of locations given predictions of where the fire may spread, and an ACTIVATE task that decides which locations are far enough from the fire zone yet still in the vicinity for evacuated residents, and enough capacity for everyone.
The ability to break down the large CHOOSE task into sub-tasks illustrates how smaller decisions help resolve the overall decision. 
This example illustrates two properties of our typology:

\smallskip\noindent \textbf{Composability:} The sub-decision tasks are interconnected and describe a flow among decisions. 

For example, CREATE decides potential options and ACTIVATE then decides which options meet given criteria.
This flow reflects how decision-making processes are structured as sequences of information transformations, supporting a process-centric view of decision-making (design goal \textbf{\ref{goal:process-centric}}). 
Moreover, the ordering of these tasks implies a sequence that captures iterative and dependent structures common in real-world decision-making (design goal \textbf{\ref{goal:iteration}}).
In the example above, the execution of CREATE must occur before the ACTIVATE decision because the information output by CREATE is the information evaluated in ACTIVATE.
The property of composability is important for describing visualization tasks in general, as demonstrated by the task typology by Brehmer et al.\cite{brehmer_multi-level_2013}.
Composability allows us to describe the relationship between decisions, which in turn helps us understand how information informs decisions (\textit{e.g.}, we know the set of options output by the ACTIVATE decision are a good set to CHOOSE from) and is transformed by decisions (\textit{e.g.}, predictions of wildfire spread into potential evacuation locations).


\smallskip\noindent \textbf{Decomposability:} The decomposition of the CHOOSE task in the above example into sub-tasks implicitly defines a hierarchical structure of decision tasks (design goal \textbf{\ref{goal:hierarchical}}).
They provide a high-level abstraction of a decision task, while detailing the step-by-step decision-making processes at lower levels.
The hierarchies indicate lower-level decisions that are needed to execute higher-level decisions due to its complexity.
By expanding complex decisions into less complex constituent decisions, we can describe situations when multiple disparate decisions make-up a larger more complex decision.

\subsection{Diagramming Decision-Making Problems}
\label{sec:diagrams}
\begin{figure}[ht!]
    \centering
  \includegraphics[width=0.80\linewidth]{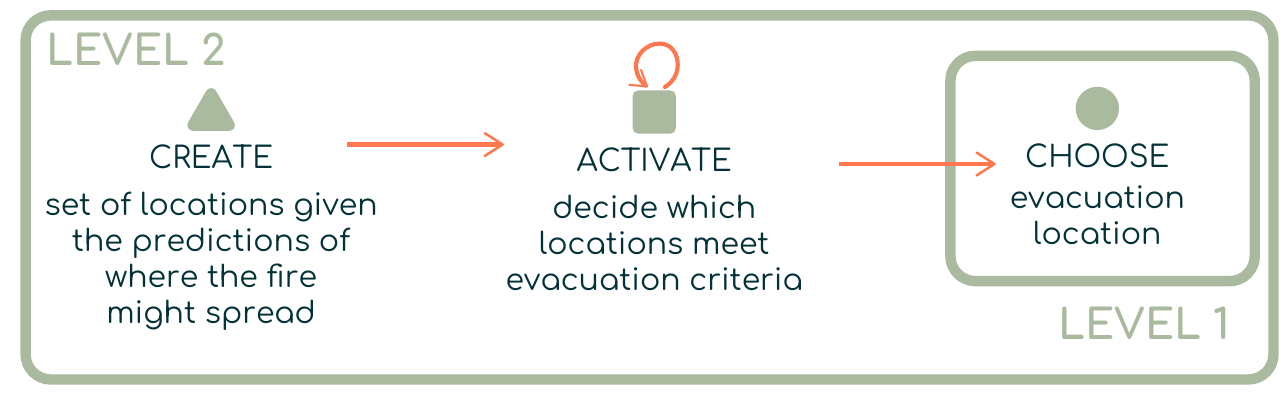 }
  \caption{A diagram of the example wildfire evacuation decision-making problem. The CREATE task has an arrow into the ACTIVATE task, indicating that the CREATE task must occur before the ACTIVATE task. The ACTIVATE task has a self-loop, which indicates the possibility that this task is executed more than once. Lastly, the ACTIVATE task has an arrow into the CHOOSE task, indicating that the locations decided on in the ACTIVATE task are the input to the CHOOSE decision.}
  \label{fig:wildfire}
\end{figure}
We use a directed graph, referred to as a ``diagram,'' to illustrate decision problems represented in our typology.
An example for the wildfire evacuation decision is shown in Figure \ref{fig:wildfire}.
In this diagram, decision tasks are depicted using distinct shapes: a circle for CHOOSE, a square for ACTIVATE, and a triangle for CREATE. These symbols are supplemented with free-form annotations in domain-specific language to provide details about the decisions being made. Arrows indicate the flow of information between decisions, with annotations on the edges specifying what information is being transferred. Additionally, the ``LEVEL X'' notation is employed within the case studies to represent the hierarchy of decisions: a LEVEL 1 decision corresponds to the overall goal, a LEVEL 2 decision supports the LEVEL 1 goal, a LEVEL 3 decision supports a LEVEL 2 decision, and so on.

\section{Case Studies}
\label{sec:case-studies}

To illustrate further our typology and its properties, we present four case studies where our typology is used to describe user decisions in the context of decision-support tools. 

\subsection{Case Study 1: Homefinder Revisited} \label{subsec:homefinder}

\begin{figure}[ht!]
    \centering
  \includegraphics[width=0.80\linewidth]{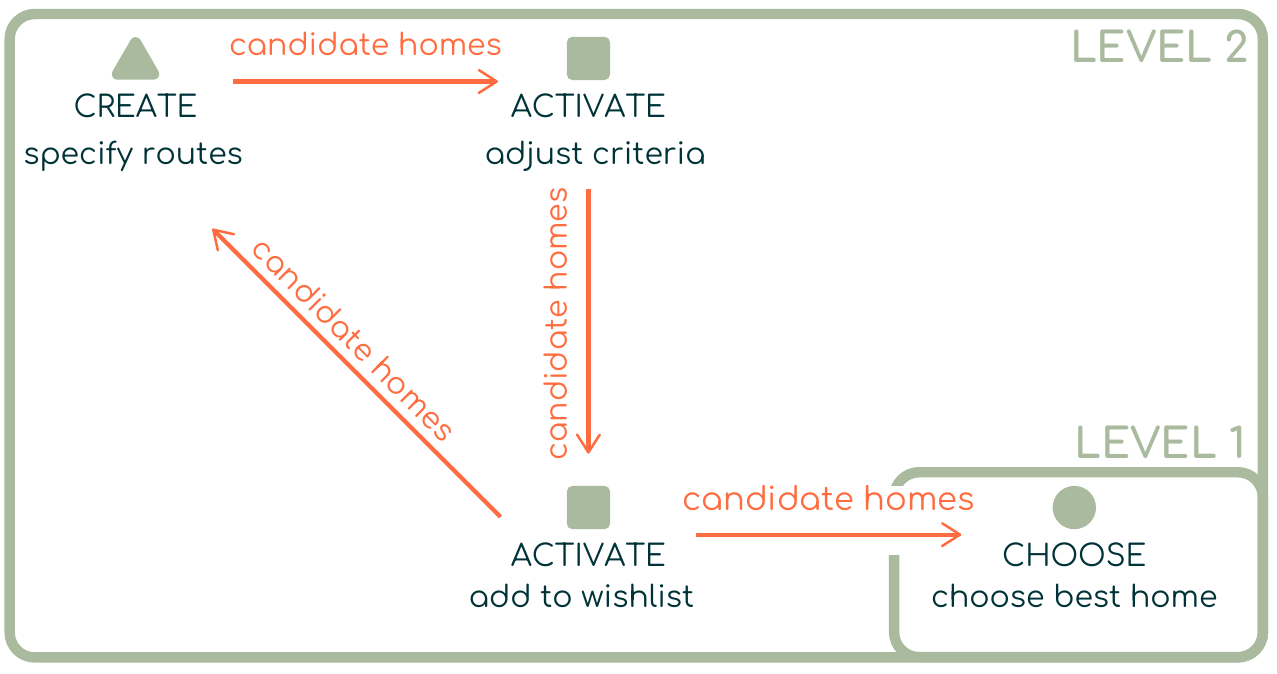}
  \caption{Homefinder Revisited\cite{weng2018homefinder} decision diagram.}
  \label{fig:homefinder}
\end{figure}

The paper ``Homefinder Revisited''\cite{weng2018homefinder} introduces ReACH (Reachability-Aided Contemporary HomeFinder), a visual analytics system aimed at aiding users in discovering their ideal homes. 
In ReACH, users input their daily routes, activities, and preferences, upon which the system generates potential homes by considering factors like reachability, proximity to amenities, and user-specified preferences. 
With its interactive interface, users can query, filter, and evaluate home candidates based on multiple criteria, ultimately choosing their ideal home.

\smallskip\noindent \textbf{Level 1:} At the root level, ReACH supports the user to CHOOSE their ideal home from a number of candidates.

\smallskip\noindent \textbf{Level 2:} Expanding on the CHOOSE decision, a user needs to first CREATE home candidates from their routes, activities, and preferences. The user performs the ACTIVATE decision to add or filter the candidates by refining the search criteria (e.g., parameters for reachability, preferences, etc.). The user then decides the homes that they would like to add to their wishlist (an ACTIVATE decision) and iterates on the process until the wishlist is complete. Finally, the user CHOOSES the ideal home from the wishlist.

\smallskip \noindent \textbf{Discussion:} The decision-making framework described in the Homefinder Revisited paper represents a classic multi-criteria decision-making (MCDM) task. 
Typically, an MCDM task follows a diagram of CREATE (creating design criteria) $\rightarrow$ ACTIVATE (filtering and refining the criteria) $\rightarrow$ CHOOSE (selecting the best options). 
ReACH adds an additional decision task to this process, allowing users to add or remove candidates from a wishlist.



\subsection{Case Study 2: Umbra} \label{subsec:umbra}
\begin{figure}[th!]
\centering  \includegraphics[width=\linewidth]{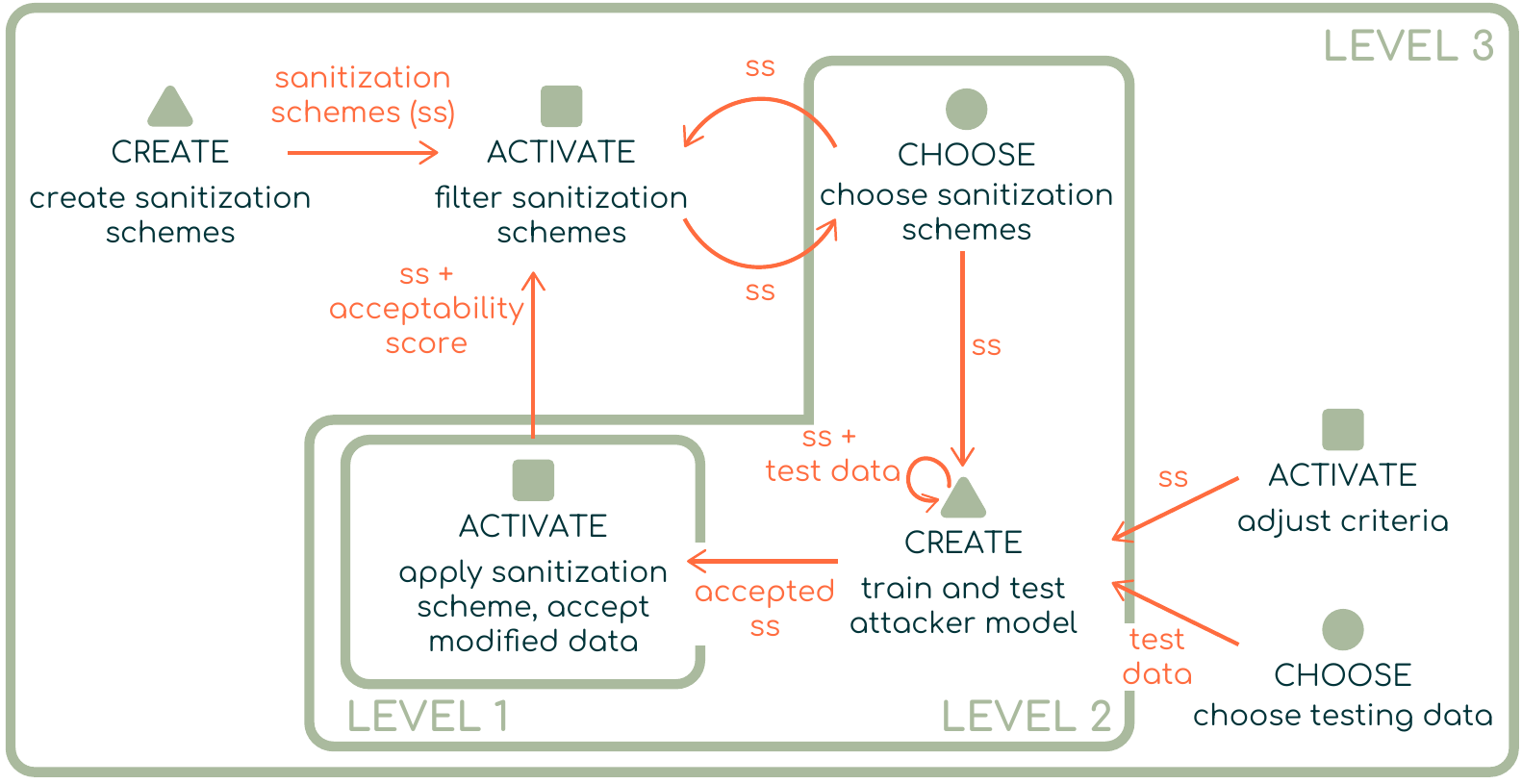}
  \caption{Umbra\cite{wang2020umbra} decision diagram}
  \label{fig:umbra}
\end{figure}
Umbra\cite{wang2020umbra} is a visual analytics tool that helps address the complex problem of sanitizing sensitive information within datasets without compromising data characteristics, particularly in scenarios involving medical data, personally identifiable information, and classified data. Umbra is semi-automated in that it can: (1) automatically identify potential risks, (2) suggest sanitization actions and schemes, and (3) perform simulated attacks. Using Umbra, an analyst would review the potential risks, run simulated attacks, adopt or develop a sanitization scheme, and finally execute the scheme on a given dataset and disseminate the result.
The decision tasks supported in Umbra can be viewed hierarchically and are illustrated in Figure \ref{fig:umbra}:


\smallskip\noindent \textbf{Level 1:} At the root level, Umbra supports the ACTIVATE decision by evaluating whether the modifications to a sanitized dataset are acceptable, based on its remaining utility and the level of privacy achieved.

\smallskip\noindent \textbf{Level 2:} At the next level, the ACTIVATE decision can be expanded to include two additional decisions: a CHOOSE decision, where the users choose what sanitization scheme to apply to the dataset and where to apply it, and a CREATE decision, where a simulated attack is generated. This forms a loop: (1) choose a sanitization scheme, (2) run the simulation, and (3) accept or reject the sanitized data.

\smallskip\noindent \textbf{Level 3:} The simple loop can be decomposed to include more nuanced decisions in the process. The CHOOSE decision of a sanitization scheme can be expanded into its own decision-flow that includes creating a sanitization approach (a CREATE decision), and deciding whether to accept the approach (an ACTIVATE decision). Similarly, the CREATE decision from Level 2 on generating a simulated attack can be decomposed further into deciding on a plan of attack (an ACTIVATE decision) and deciding whether to use custom or default test data (a CHOOSE decision). 

\smallskip \noindent \textbf{Discussion:} 
In the first case study, the ``Homefinder Revisited'' paper describes the fundamental user task at the root level as a CHOOSE decision for choosing an ideal home from a finite list of candidates. 
However, in this second case study, the number of candidates is potentially infinite, as Umbra can generate as many sanitization schemes as required.
As a result, the Umbra system is designed to assist users in examining and refining a scheme, culminating in the decision of whether the scheme can be ACTIVATED and applied to sanitize a dataset.


\subsection{Case Study 3: Centaurus}
\label{subsec:centaurus}

\begin{figure}[ht!]
\centering\includegraphics[width=0.80\linewidth]{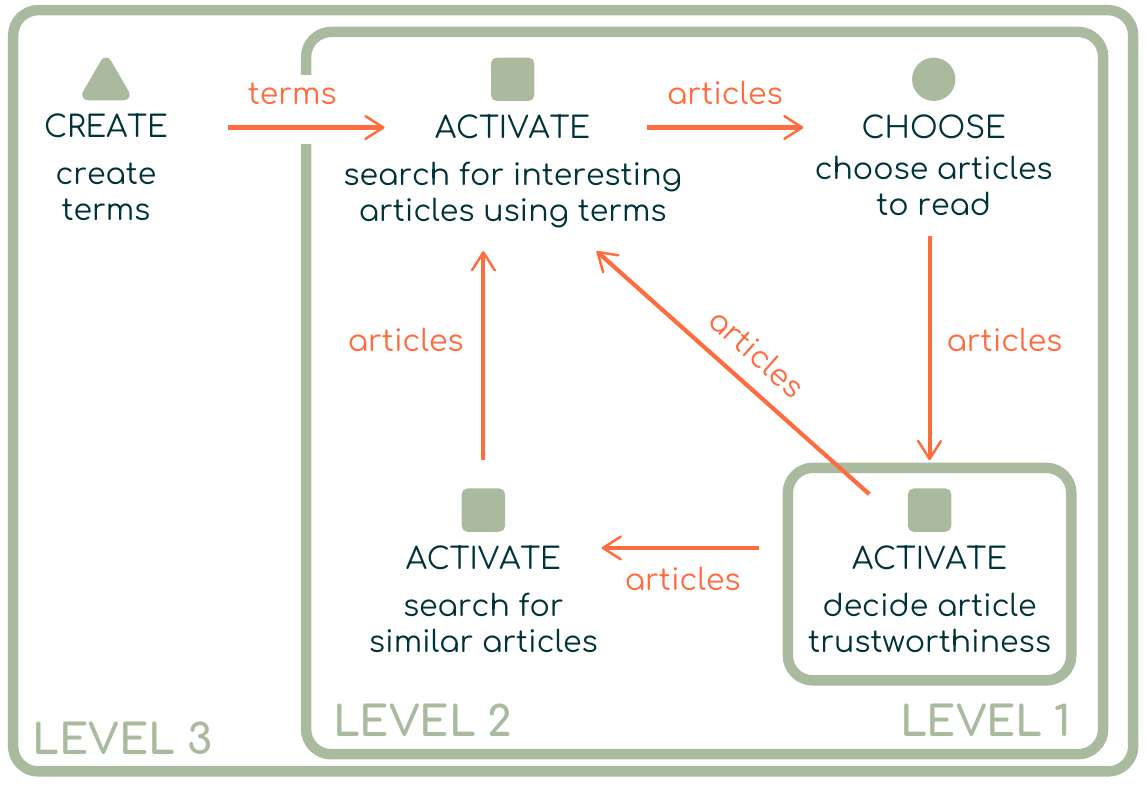}
  \caption{Centaurus\cite{dowling2020semantic} decision diagram.}
  \label{fig:centaurus}
\end{figure}

The Centaurus tool was utilized in a case study to evaluate the truthfulness of claims by analyzing related news articles\cite{dowling2020semantic}. Users were presented with a claim, such as a tweet, accompanied by a mix of trustworthy and untrustworthy news articles. Through Centaurus, users engaged with these articles to extract relevant information and assess the trustworthiness of each article. The primary focus was on understanding how users utilized Centaurus to formulate their final evaluations, irrespective of their stance on the original claim's truthfulness.
Figure \ref{fig:centaurus} illustrates the following breakdown of the decision-making tasks:

\smallskip\noindent \textbf{Level 1:} The core of the Centaurus decision problem is an ACTIVATE decision, wherein users evaluate the trustworthiness of articles.

\smallskip\noindent \textbf{Level 2:} The Level 1 ACTIVATE decision is supported by additional decisions -- an ACTIVATE to select relevant articles based on search terms, a CHOOSE to choose articles for deeper analysis, and another ACTIVATE to search for similar articles after a user has made a decision about a potential article to investigate further. 

\smallskip\noindent \textbf{Level 3:}
The ACTIVATE decision for selecting articles based on search terms is expanded to include a CREATE node where a user can manually enter keywords as search terms.

\smallskip \noindent \textbf{Discussion:}
While the domains of Centaurus (determining article trustworthiness) and Umbra (choosing data sanitization) may seem different, they exhibit notable similarities in the decision-making processes. Figure~\ref{fig:centaurus} shows the decision diagram of Centaurus that uses a similar layout as the Umbra diagram in Figure~\ref{fig:umbra}. 
By comparing the two, one can observe that Centaurus features the same three tasks at the top row as Umbra (CREATE, ACTIVATE, CHOOSE), and the bottom row also shares similarities. This comparison highlights the utility of our typology in evaluating and contrasting the decision tasks of seemingly disparate decision-support visualization tools. 



\subsection{Case Study 4: WireVis}
\label{subsec:wirevis}
\begin{figure}[ht!]
  \centering\includegraphics[width=0.95\linewidth]{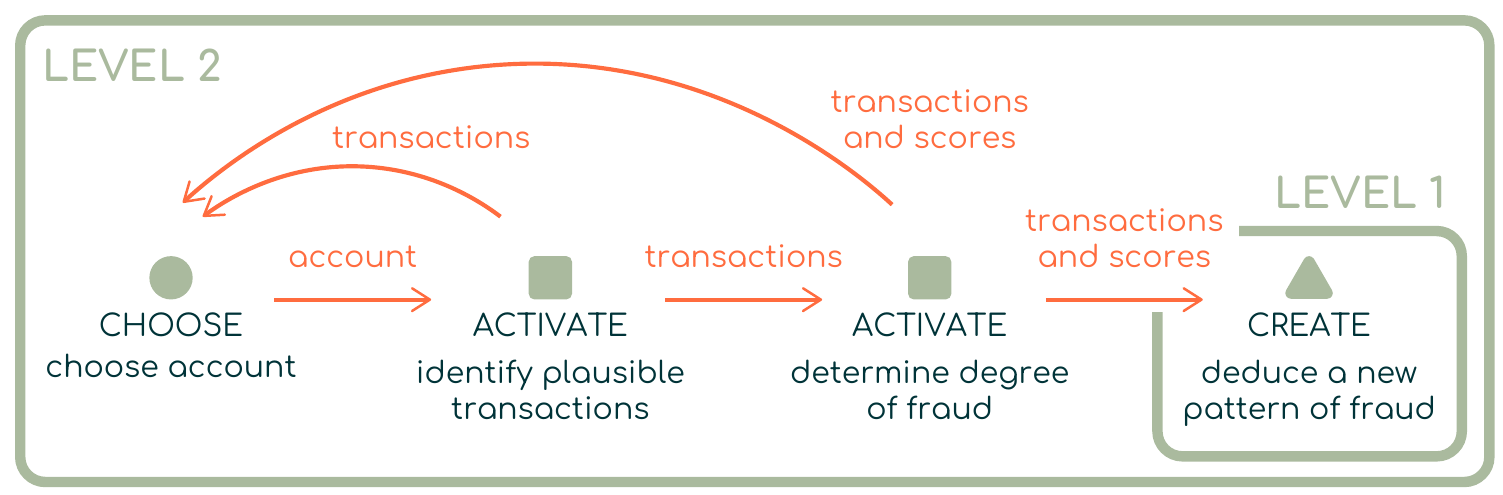}
  \caption{WireVis\cite{chang2007wirevis} decision diagram.}
  \label{fig:wirevis}
\end{figure}

The WireVis paper introduces a visual analytics system designed for the analysis of large amounts of financial wire transactions, aiming to uncover suspicious or fraudulent activities such as money laundering\cite{chang2007wirevis, chang2008scalable}.
The use of the WireVis system entails two distinct yet interconnected decision tasks: (1) identify fraudulent transactions and their associated accounts, and (2) discover previously unknown strategies used to commit financial fraud.
In both cases, analysts use a coordinated set of visualizations to examine financial transactions from multiple perspectives, including keywords, amounts, time, and account relations. 
The system can automatically generate hierarchical clusters and perform pattern matching, thereby assisting analysts in their investigations.

Focusing on the task of ``discovering previously unknown strategies used to commit financial fraud,'' the decision tasks can be represented as:

\smallskip\noindent \textbf{Level 1:} The core objective can be described as a CREATE decision where an analyst aims to synthesize and discover a previously unknown ``financial fraud strategy'' based on suspicious financial transactions and activities.

\smallskip\noindent \textbf{Level 2:} In order to CREATE the strategy, an analyst first need to CHOOSE the most promising accounts to investigate. For these accounts, the analyst determines the set of their transactions that might exhibit an unusual pattern of behavior (an ACTIVATE decision). These transactions are examined based on a list of criteria, such as the amount, the keyword used, the date patterns, etc. (an ACTIVATE decision). If deemed suspicious, an analyst might be able to deduce a previously unknown strategy of financial fraud (a CREATE decision). 

\smallskip \noindent \textbf{Discussion:}
The WireVis case study diverges from previous examples in that the root decision is not to simply choose or filter out potential options. Instead, it involves the creation of insights assembled from evidence gathered in the tool. 
Interestingly, the CREATE decision while important in the decision-making problem addressed by this tool, is not directly supported by the visualization system.

\section{Evaluation}
\label{sec:evaluation}



\begin{figure}
    \centering
    \includegraphics[width=\linewidth]{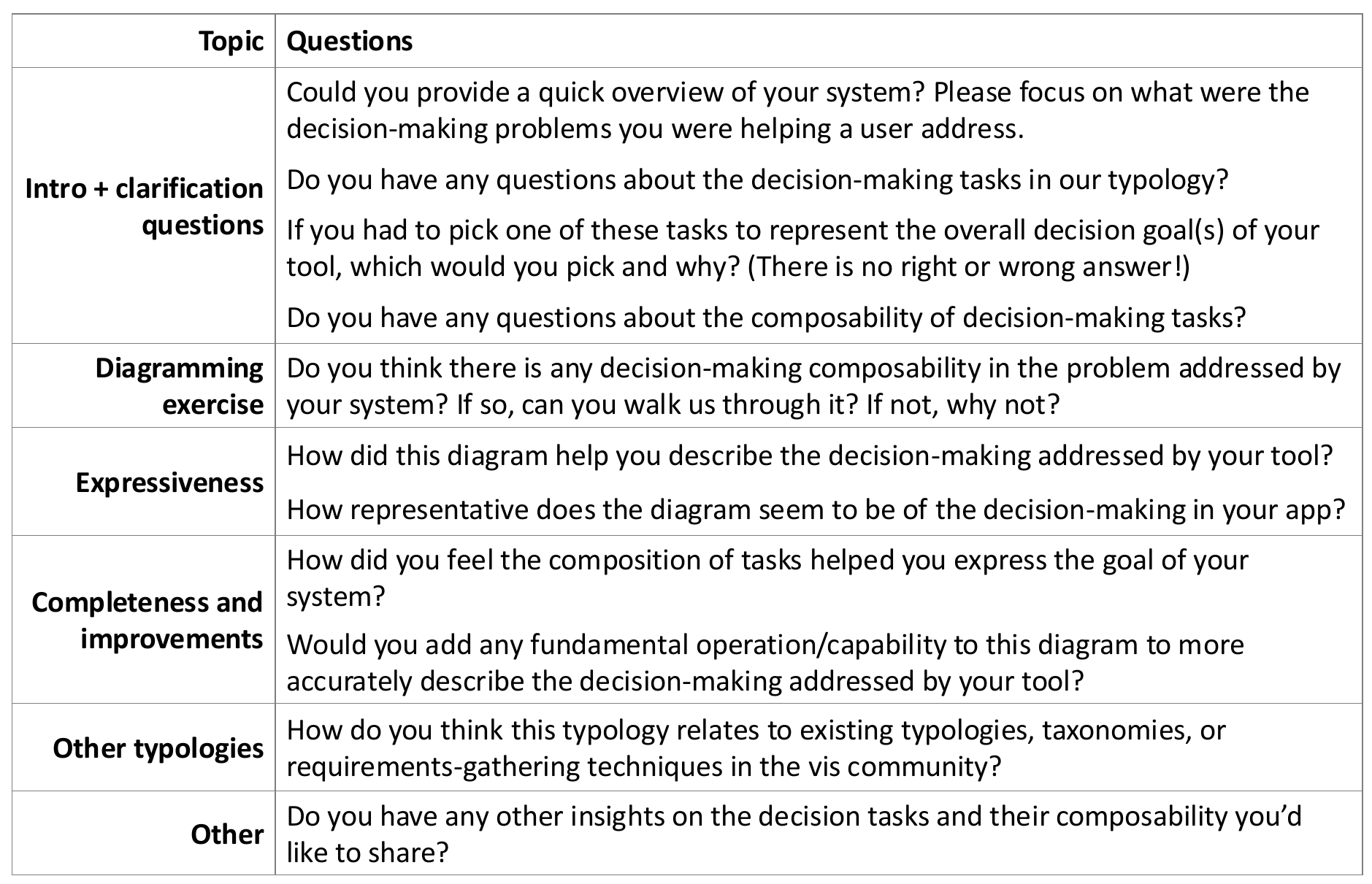}
    \caption{Interview questions we asked our participants. 
    }
    \label{fig:interview-1-questions-table}
\end{figure}

To further validate our typology, we recruited nine visualization experts who built decision-support tools in the past to evaluate our typology and identify its strengths and weaknesses. 
Three of our participants are assistant professors, four are scientists at national laboratories, one is a senior visualization developer at a market research company, and one is a PhD student in visualization. 
The interviews were semi-structured, guided by a set of pre-determined open-ended questions (please see figure \ref{fig:interview-1-questions-table}) while allowing room for deeper exploration of particular topics as they emerged.
Specifically, our questions examined three dimensions proposed by Beaudouin-Lafon\cite{beaudouin2004designing}: (1) \textit{descriptive power}: ``the ability to describe a significant range of existing [designs]'', (2) \textit{evaluative power}: ``the ability to help assess multiple design alternatives'', and (3) \textit{generative power}: ``the ability to help designers create new designs.''
Each interview lasted approximately one hour and was recorded with the participants' consent.

During each session, we introduced the typology and its properties using the car buying example, as detailed in Sections \ref{sec:typology} and \ref{sec:composability}. 
We encouraged participants to ask questions and seek clarifications to ensure they fully understood the typology and its properties. 
Following this, we conducted a collaborative sketching session using shared presentation slides, where participants used the typology to describe the decision-making tasks supported by a visualization-based decision-support tool they had designed in the past. These tools varied widely in purpose, supporting tasks such as detecting fraud in financial transactions, selecting items to purchase, generating and testing data-driven hypotheses, planning urban infrastructure, and determining new bus routes, among others.
As participants articulated the decision-making tasks and processes their tool supported, we collaboratively sketched corresponding diagrams (see Section \ref{sec:diagrams}). Participants were then invited to review and refine these diagrams until they accurately reflected the decision flow experienced by their users. To preserve participant anonymity, we have deliberately omitted references to tools with associated publications. The remaining tools were developed in industry or are protected under non-disclosure agreements, and thus cannot be disclosed. Slides and finalized diagrams from the interviews are available in the Supplemental Material.

At the conclusion of each interview, we invited participants to provide additional feedback on the criteria used by Brehmer et al.\cite{brehmer_multi-level_2013} when evaluating their typology: correctness (subsection \ref{eval-correctness}), utility in the design process (subsection \ref{eval-utility}), and comparison to other visualization typologies (subsection \ref{eval-typologies}).
We also asked them to share suggestions for improving the typology both during the interview and afterward via email if they had further insights. 
We discuss the limitations of the typology gathered from these interviews in more depth in Section \ref{sec:limitations}.


\subsection{Correctness}
\label{eval-correctness}

The participants discussed two topics related to the correctness of the typology: (1) how \textbf{complete} is the typology, and (2) how well the participants were able to use the typology to \textbf{express} their decision-making problems.

\smallskip \noindent \textit{Completeness}

In evaluating the correctness of our typology of decision-making tasks for visualization, we first assessed its completeness to determine if there were any missing decision tasks in our typology that could limit its ability to capture all aspects of decision-making. 
We adopted Gleicher et al.'s definition of completeness by replacing ``designs'' with ``decision-making problems'', that is, the typology needs to cover all decision-making problems we have encountered \cite{gleicher_visual_2011}.
The participants in our study found that our typology was indeed comprehensive, allowing them to represent all decisions within the problem domain. As noted by one of the participants: ``\textit{I think [the typology] is a pretty good set of high level operations. [...] It took me a little bit to really clarify the distinction between activate and choose. But once I got there, it makes a lot of sense.  I don't know if I can really think of another type of operation that I felt was missing. It seemed, at least in the case of my problem, a good set [of operations] to describe what's going on}.'' 
Other participants expressed similar sentiments.
Overall, all of the participants were able to describe their decision-making problems using our typology, and none of them identified any missing decision tasks or fundamental operations in our typology.

Interestingly, a few participants incorporated additional annotations when describing their decision-support tools using our typology.
These annotations were used to detail the contexts of these participants' decision-making problems, including temporal information (e.g., time duration for how long a task would take or how frequently a task needs to be performed), roles (e.g., who in an organizational hierarchy should make the decision, or in a collaborative setting, which collaborator is responsible for which task), and decision styles (e.g., rational, conceptual, analytical, etc.).
Such contextual annotations complement our typology and suggest an interesting future extension of our work.
We explore these concepts further in Sections \ref{sec:limitations} and \ref{sec:future-work}.


\smallskip \noindent \textit{Expressiveness}


During this study, it became clear that our typology is also expressive, meaning it can be easily used to describe decision-making scenarios.
All participants expressed their proficiency in composing and decomposing decisions with the typology, enabling them to construct cohesive representations of the problems their respective visualization tools addressed. They noted that the typology captured the decision-making processes inherent in problem domains and facilitated the visualization of user interactions with decision-support tools. 
As one of the participants said, ``\textit{[The use of the typology] really kind of helped highlight some of the complexity around the design goals that went into it. And it's definitely reflective of conversations that [my team and I] had going into the design of the tool and sort of throughout its life span. So, in that sense, I think it's very representative.}''

This sentiment is perhaps best summarized by a participant: ``\textit{Breaking down this higher level activate task...in terms of how [the user] actually interacts with the system and what the system is doing for them, I think [the typology] helped tremendously. [...] I was able to describe step by step what people were doing in the system, and those steps mapped really cleanly to the [three decision-making] tasks.}''

\subsection{Utility in the design process}
\label{eval-utility}
During the interviews, the participants discussed the potential applications of the typology. We categorize their feedback into three themes: the use of the topology for (1) \textbf{analysis and reflection} for guiding the design process of visualization tools, (2) facilitating \textbf{communication} between developers, domain experts, and end-users, and (3) \textbf{categorizing and organizing} VIS publications related to decision-making.

\smallskip \noindent \textit{Analysis and Reflection of Design}


Participants reported that structuring decisions using our typology diagram helped them break down decision problems, identify potential decisions that were not initially considered, and recognize what decisions lacked support in their system. 
In particular, the typology diagram served as an externalization of mental models for the participants to better reflect on the overall decision-making process, allowing them to step back from the development of tools and consider the bigger picture. 
As explained by one of the participants, ``\textit{I think this [exercise using the typology] was actually very valuable from my standpoint in reflecting on what we are building. Take a step back and reflecting on how these decisions, what we understand, and from an abstraction and typology standpoint, what they mean. So it's been actually helpful in that regard}.''

Another participant further delineated the benefits of using the typology, ``\textit{I see this as a way [...] for us to understand clearly what we are doing. Not to focus a lot on \textbf{how} we do it, but \textbf{what} we want to do. [...] Breaking down the task like that could help me to also think about new possibilities for decision-making that I didn't consider before or didn't have time to.}''




\smallskip \noindent \textit{Facilitate Communication}

During our interviews, participants highlighted the typology's potential to facilitate communication with domain experts and end-users during the design process by providing a higher-level perspective. 
One participant articulated, ``\textit{If I had this taxonomy like in your work, I can better communicate with the experts. [...] I think with this taxonomy maybe we can begin from a more high level perspective, like what does the system look like or how this system would adjust its requirements from a higher perspective. And then we can drill down to each part, for creation, for activation, for choose. What kind of interface we can design for them. I think [this diagram] would facilitate this kind of brainstorming sessions.}''

By facilitating better communication, a participant believed that the use of the typology could improve shared understanding of the design process between the stakeholders, thereby improving the final design. ``\textit{I think having some kind of decision-making type of language might help bridge that gap between} `I as the developer, designer, researcher, I think I know what you mean, I think I know what you want, and what your processes are' \textit{as opposed to the user being like} `I don't even know how to better explain this.'''
Similarly, another participant compared the use of the typology to the standard practice that relies on a list of design requirements in the design process and noted, ``\textit{I think this [diagram] is a better specification of the process than the list of requirements, for sure, because at least I can see here the process of the decision, not a well laid set of requirements.}''

Although our evaluation did not involve domain experts or end-users, our participants speculated that our typology would be intuitive for their domain experts and end-users to understand, thereby improving the design process.
One participant noted, "\textit{I don't think you could really present [Brehmer et al.'s task typology] to someone outside of the [visualization] community and explain it in 10 minutes and they could like really, intuitively, get it. [Your typology] feels much more accessible.}"
Another participant said, ``\textit{I haven't practically used the diagram [with a domain expert, but] I think the expert could provide feedback directly on the diagram. Basically, if they want more alternatives, or if they think that the current way of generating alternatives is not satisfactory, they can request to add edge backward to the creation tasks. So our system design will improve.}'' This direct engagement with the diagram has the potential to improve the collaborative refinement of system design, allowing experts to contribute insights that could enhance decision-making capabilities.

\smallskip \noindent \textit{Categorization and Evaluation of Publications}

Some participants highlighted the potential of our typology in aiding the research community in categorizing and organizing publications relating to decision-making. In our interviews, three participants utilized our typology to sketch diagrams representing different versions of their tools or extensions published in subsequent papers, demonstrating its capacity to compare decision processes across various tools. Through these diagrams, they were able to discuss the nuances and distinctions between these tools, highlighting areas with varying levels of decision support. This process enabled them to evaluate and compare their own tools' effectiveness.

Some participants expressed that our typology could serve as a valuable tool for evaluating decision support tools due to its ability to organize the decision process into distinct parts and delineate where decisions begin and end. One participant noted, ``\textit{In my experience as a reviewer, I don't explicitly ask authors to follow specific patterns or taxonomies to organize their work. However, if they use a taxonomy, it will be easier to follow their paper. Basically, they can organize decision-making papers into these few stages, making it easier to evaluate this type of work}.''


Instead of delving into the minutiae of individual visualizations in the review process, one participant suggested that our typology could offer a framework to assess systems holistically, stating that, ``\textit{I think [your typology] could be helpful as well when we evaluate visualizations. [In reviewing visualizations] we go very low level [evaluating if] people [can] read a certain value [from a visualization]. [...] Thinking about [a decision support] system [using your typology], like in this way, might be also helpful to evaluate the systems overall, because [as a reviewer] I know where I start and where I want to end. And I can see how [a user would] follow these steps to get to the final process.}''

\subsection{Comparison with Existing Task Taxonomies}
\label{eval-typologies}
All of our participants are researchers in the VIS community who are well-versed in the VIS literature.
Many drew comparisons between our typology and existing task taxonomies, such as those proposed by Amar et al.\cite{amar2005low} and Brehmer et al.\cite{brehmer_multi-level_2013}.
In this section, we summarize their comments on how our typology compares to existing task taxonomies in terms of: (1) the \textbf{level of task abstraction} and (2) its \textbf{flexibility} in practical use.

\smallskip \noindent \textit{Level of Task Abstraction}


During discussions, participants noted that our typology is more oriented towards modeling processes in contrast to the focus on data, analytical operations, and interactions emphasized by existing task taxonomies. This approach resonated with them as they considered it more suitable for describing complex decision-making tasks. One participant articulated this sentiment as follows: ``\textit{I think [existing task taxonomies] can be suitable for all visual analytics systems and it’s a bit low level [...]. [But] if you want to describe my system with this type of taxonomies you have to go very low level, like presenting values or like sorting the alternatives. [...] It’s a bit cumbersome to describe the system using this taxonomy. Yeah, yours are better. I think your taxonomy is doing a better job at describing decision-making tasks or decision-making systems, because it’s from the number of alternative perspective. So I think that the taxonomy more fits in a decision-making scenario.}''

Other participants also noted the challenge of using existing task taxonomies to describe a complex design-support visualization system and suggested that a higher-level typology is better suited for describing the user's needs and perspectives.
As stated by one of the participants, ``\textit{[Using existing taxonomies] is very low level to understand a system like [my system]. For the final users, the user is not thinking about reading this bar chart. [The] users say} `can I choose my actual scenario?'''
Another participant agreed, "\textit{[Your typology] is abstract, but it helps you think about the decision process. This is closer to the user than the \textbf{why} of a visualization [in the typology by Brehmer et al.], because you don't specify, for example,} 'I want to present data,' \textit{or} 'I want to discover data.' \textit{[...] [With a] diagram [using your typology], [...] I can see a process that I can match well with what we did in [our system].}''

\smallskip \noindent \textit{Flexibility to Data and Context}

The participants in our evaluation highlighted the flexibility of our typology, particularly in its adaptability to different data and contexts while maintaining well-specified functions. 
The participants could compose and decompose decisions according to their decision tasks and preferences for modeling the problems. 
For example, one participant chose to represent their tool's addressed problem hierarchically, noting, ``\textit{This does feel more hierarchical. I'm definitely reminded of task abstraction type of work, which tends to go a step beyond just requirements gathering.}''

Other participants opted for the diagram representation, utilizing cycles and arrows to depict iterations within their decision-making process. This approach allowed for clear visualization of feedback loops, with decisions feeding back into the initial stages of the process. 
One participant who adopted this method remarked, "\textit{[Brehmer et al.'s typology] very much still had kind of a rigid application to it, whereas I think the typology that you’re proposing feels like it can be more flexible.}"


The general agreement among participants was that our typology stands out for its flexibility, allowing for diverse representations of decision-making processes tailored to specific needs and contexts. A participant summarized this quite succinctly by stating, ``\textit{I like that a lot as part of your typology, is that it feels simple and flexible, yet powerful.}''
\section{Design Implications}
\label{sec:design-implications}

The evaluation presented in Section \ref{sec:evaluation} examines the utility of our typology to describe decision-making problems supported by visualization systems. 
Here we present initial findings on design implications of our typology gathered from follow-up discussions with available interview participants.




Two of our nine interview participants, a PhD student specializing in design studies (P1) and a visualization developer from industry (P2), were available for a follow-up discussion about their visualization tools. 
We presented them with the decision-making diagrams they had created during the initial evaluation of our typology. 
During this session, we asked them to reflect on the diagrams and sketch potential design changes informed by the insights gained. 
Specifically, they were asked to: 1) identify which decision tasks were well-supported and which lacked adequate support, 2) sketch design improvements for areas where the system did not provide sufficient decision support, and 3) conceptualize a new design from scratch based on the problem outlined in the diagram. 
Both participants utilized our typology to enhance their designs, incorporating new views, layouts, and interactive features to better support and expand the decisions and processes within their tools.



\begin{figure}[t!]
    \centering
    \includegraphics[width=\linewidth]{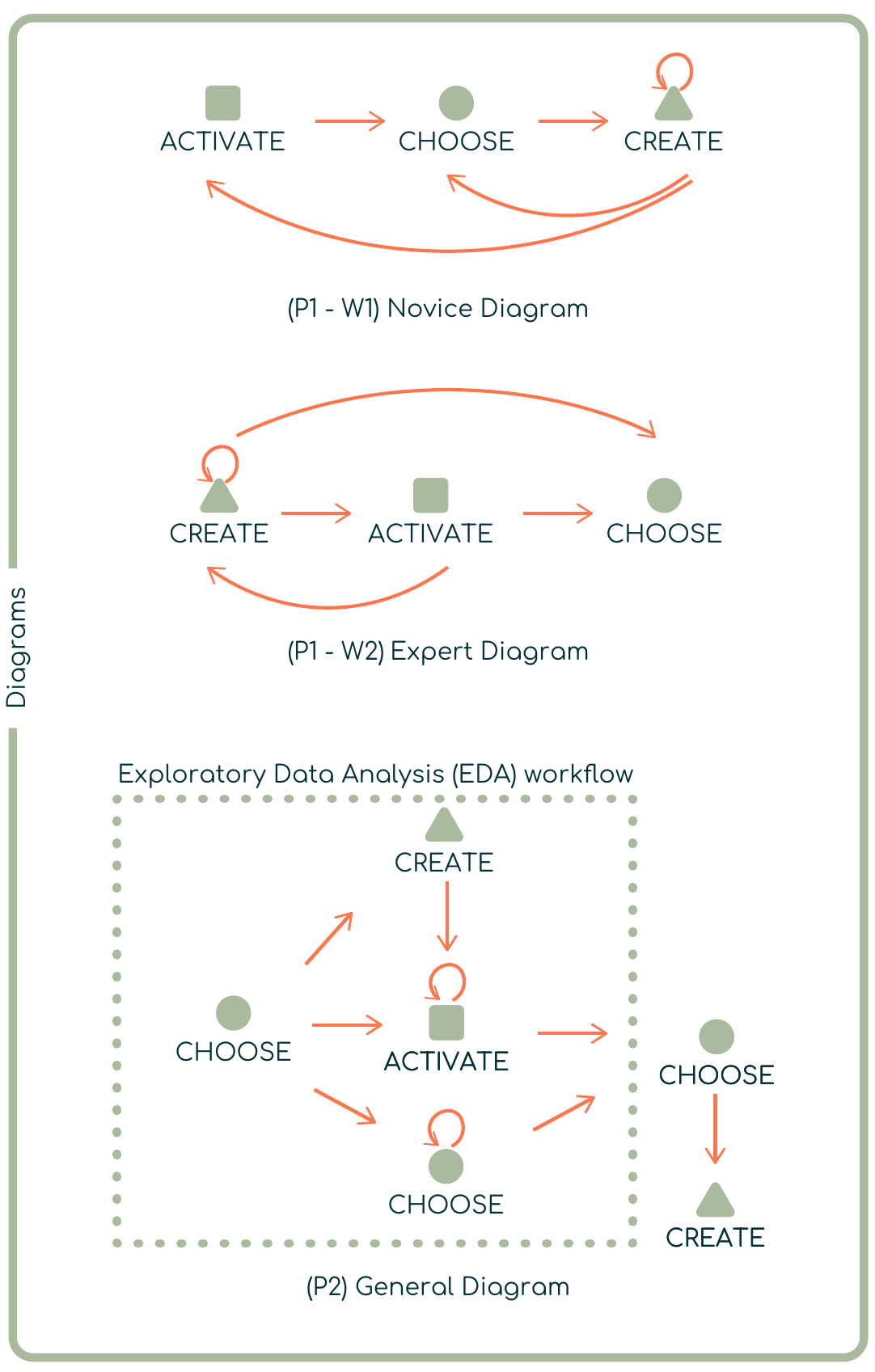}
    \caption{Typology diagrams elicited in the first interview. The drawings have been recreated to match the paper's theme, but the content remains true to what the participant sketched during the interview.}
    \label{fig:redesign-diagrams}
\end{figure}
\begin{figure}[t!]
    \centering
    \includegraphics[width=\linewidth]{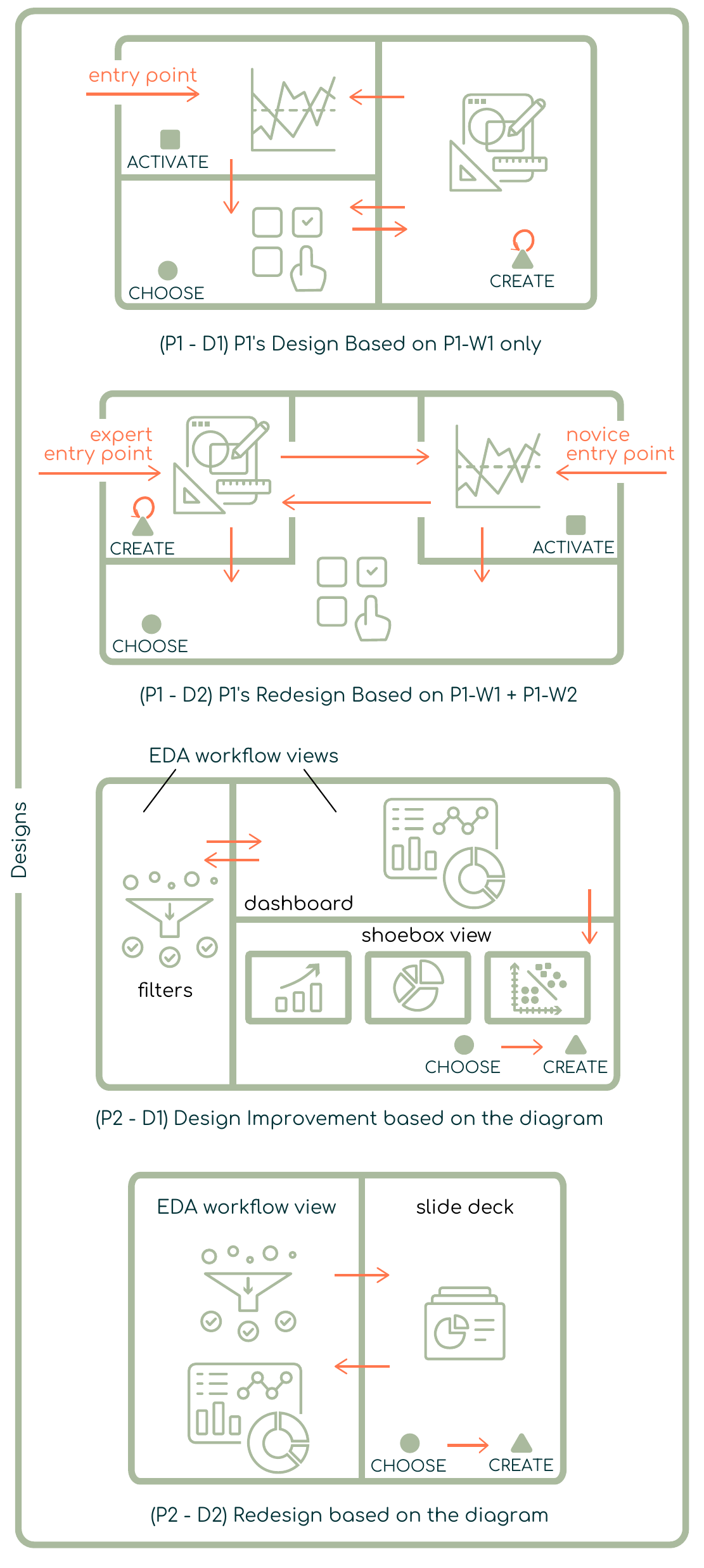}
    \caption{Design improvements and redesigns sketched by participants during the preliminary study. Icons representing visualizations have been added to indicate coordinated views, but these were not part of the participants' original sketches.}
    \label{fig:redesign-designs}
\end{figure}

Using the diagram created in our previous interview, P1 redesigned their tool to better accommodate both novice and expert users.
The initial tool design had a workflow where numerous eligibility criteria are evaluated, filtered, and ultimately selected, leading to the creation of more restrictive criteria that feed back into the decision-making process (Figure \ref{fig:redesign-diagrams} P1 - W1).
This approach catered to novice users lacking the experience necessary to create specific criteria without overlooking crucial variables.
However, after creating the decision-making diagram in our first interview, P1 realized that expert users required a different workflow that allows them to create criteria from scratch, exhibited by the second diagram (Figure \ref{fig:redesign-diagrams} P1 - W2) drawn in the follow-up discussion.
In the initial tool design (Figure \ref{fig:redesign-designs} P1 - D1), only one of these workflows is supported at a time.
In redesigning the tool, the participant integrated both workflows by allowing two separate entry points into the tool (Figure \ref{fig:redesign-designs} P1 - D2).
This redesign supports a broader range of users and allows flexibility in utilizing both workflows when needed.
Furthermore, P1 used the decision-making diagrams to reflect further on the design and interaction across the views associated with each decision.
This resulted in P1 adding coordinated views to facilitate the ACTIVATE and CREATE decisions, and replacing an overview visualization with more interactive capabilities to support experts in the CREATE decision. Thus, the typology diagram might inform high level design patterns, such as adding coordination between views that correspond to decisions that are linked in the diagram, or considering the information needed on the landing page of the system depending on which is the first decision that needs to be made. However, P1 hinted at the need for some correspondence from the decision-making typology and design decisions, e.g. the choice of certain visual encodings. Exploring the mapping between decision-making tasks and design patterns is an important direction for future work.

P2 reflected on the lack of support in their visualization tool for some of the decisions contained in their decision diagram.
The tool primarily supports the exploratory data analysis stage (P2's EDA workflow in Figure~\ref{fig:redesign-diagrams}) where users iteratively explore data and make intertwined decisions, followed by a sequential stage that culminates in a ``CREATE report'' decision (the CHOOSE $\rightarrow$ CREATE motif in P2's diagram in Figure~\ref{fig:redesign-diagrams}).
P2 noted that the final CREATE decision is not well-supported by the tool, primarily because the initial requirements gathering focused more on the iterative, exploratory, and data-driven aspects of the process, rather than the sequential stages of decision-making. As a result, tasks like creating presentation slides are left to existing tools rather than being integrated into the system.
However, P2 noted that a possible improvement in their tool that allows users to save intermediate results during exploration could better support the final CREATE decision.
To address this, P2 sketched and considered two possible changes in the tool design: 1) a ``shoebox'' view where users drag and drop elements and insights gathered during the EDA process (Figure \ref{fig:redesign-designs} P2 - D1), and 2) a split view with the ability to drag and drop results directly into presentation slides (Figure \ref{fig:redesign-designs} P2 - D2)).
P2 ultimately preferred the split view design as it provides continuous visibility and directly supports the final CREATE decision. Observing the differences between the design improvement exercise and the redesign exercise, we found the participant shifting from a bottom-up towards a top-down approach.
Rather than focusing on the data or individual steps of the EDA process, more time was spent considering the overall decision goal and information needs of the user to achieve that goal.

While these are only preliminary findings, there is evidence that our typology aids in the design of visualization systems.
In future work, we intend to investigate further how our typology assists in the design of decision-support visualization systems.

\section{Limitations}
\label{sec:limitations}
A few limitations of our typology emerged during our interview study: (1) the imprecise description of decision-making context, (2) the flexibility in diagram decomposition, and (3) the lack of guidance for final design decisions.

First, the typology describes decision-making tasks, but it does not account for broader contexts in which these tasks can occur. During the interview study, several participants supplemented their diagrams with annotations to provide context, without altering the core tasks or their definitions. These annotations included details such as the frequency of decision-making instances, roles within collaborative or organizational settings, and decision-making styles. While these annotations do not change the core definitions of the tasks, they often influence the design of decision-support systems.
For example, one participant suggested using the typology to depict decision-making based on different "personas," resulting in diagrams that reflect distinct roles within an organization, while another participant proposed annotating diagrams to differentiate between tasks performed by humans and those involving computer assistance or collaboration.

Second, the typology's flexibility in decomposing decision-making tasks also presents a challenge. There are no rules or exact guidance on the appropriate level of granularity to represent decision-making problems. Our participants annotated the inputs and outputs of each decision task on the arrows that connect them. However, across participants, we have seen that the level of detail varied significantly--some provided high-level information, while others included very specific dataset details. This lack of a stopping point in the recursive decomposability of the tasks mirrors limitations seen in other typologies, such as the under specified abstraction levels found in Brehmer et al.'s typology \cite{brehmer_multi-level_2013}. While this flexibility can ease diagram creation, it presents challenges in maintaining consistency.

Lastly, we primarily focused on the descriptiveness of decision-making problems, and how to use these descriptions as tools in visualization design.
However, our typology offers no direct mechanism to compare designs that satisfy the same decision-making diagram.
We acknowledge that selecting between design alternatives is influenced by various factors beyond the typology itself, such as the designer's experience, time constraints, and the decision-making style of the domain expert, which are not captured by our typology.

While our typology offers valuable insights and flexibility in describing decision-making processes, these limitations indicate areas for refinement and further exploration. In the following section, we outline our future work, which will address some of these limitations and expand upon our typology's potential applications.
\section{Future Work}
\label{sec:future-work}
While the limitations section outlined some constraints of our typology, future work will focus on leveraging the typology as a foundation for building new visualization knowledge, applications, and tools. The following points summarize our intended directions:

\smallskip\noindent \textbf{Full-Scale Design Evaluation.} A comprehensive design evaluation with designers and their domain experts will be conducted to explore the typology's utility in end-to-end visualization design settings. This will involve longitudinal studies rather than brief interviews, allowing for deeper insights into how the typology can be applied in real-world design processes.


\smallskip\noindent \textbf{Expanding Design Integration.} Informed by our user studies, future work will involve integrating the diagrams generated by our typology with corresponding visualization designs, potentially leading to the development of recommendation systems for decision-support tools, such as the Draco system \cite{moritz2018formalizing}. We plan to conduct an in-depth analysis of recurring decision-making patterns, such as multi-criteria decision-making, which could evolve into an open-source project where contributors share typology diagrams. This would enable the study of decision-making patterns and the recommendation of similar problems and design solutions.

\noindent \textbf{Connecting Decision-Making and Analytical Tasks.} Another avenue for future work is exploring connections between decision-making tasks in our typology and established analytical tasks, such as those in Amar and Eagan's taxonomy \cite{amar2005low} or Brehmer et al.'s typology \cite{brehmer_multi-level_2013}. Identifying patterns in these connections could help predict decision-making tasks based on user interaction data or suggest relevant analytical tasks based on decision goals. This would have a direct impact on the reasoning about decision-making tasks in the context of visual analytics, as it can enhance the alignment between decision support systems and the analytical workflows that users engage with, ultimately improving the efficiency of visual analytics tools. Recent work on modeling user interactions in large-scale visualization systems \cite{benvenuti2023modeling} provides a promising foundation for this direction, as it demonstrates how structured representations of user interactions (e.g., via augmented statecharts) can be used to understand and optimize user behavior. Integrating such models with decision-making task analysis could enable systems that adapt not only to performance needs but also to users’ decision intents further enhancing the effectiveness of visual analytics tools.

\noindent \textbf{Human-Machine Decision-Making.} We plan to use our typology to explore and reason about the division between human and machine decision-making in systems, and how designs can reflect or leverage this distinction.
We believe our typology can provide insights into human-machine collaboration, the effects of automating parts of the decision-making process, and the role of human intervention in AI-driven systems. This is especially crucial in visual analytics, where effective collaboration between humans and AI ensures interpretability, trust, and control over complex decision-support tools~\cite{monadjemi2023human}.

Overall, the next steps will involve applying our typology to the development and improvement of tools and decision-making processes, seeking to validate its relevance and effectiveness in real-world design and decision-making scenarios.

\section{Conclusion}

This paper presents a typology for decision-making tasks in visualization, addressing the limitations of existing task taxonomies. 
Built upon prior research and informed by design goals derived from a thorough literature review, the typology comprises three decision tasks: CHOOSE, ACTIVATE, and CREATE. 
These tasks allow for the representation of complex decision-making structures, as they can be composed or decomposed into other tasks. 
The typology demonstrates completeness, expressiveness, and utility across various decision-making contexts and is validated through semi-structured interviews with visualization experts.
During the interviews, the visualization experts expressed that the typology has the potential to be used in analyzing designs of decision-support visualization systems, facilitating communication between designers and domain users, and categorizing visualization publications on decision-making.
By providing a structured approach to understanding decision-making tasks, the typology offers a tool for researchers and practitioners alike, with implications for improving designs of decision-support tools in the future.

\bibliographystyle{IEEEtran}
\bibliography{main}

\vskip -3\baselineskip plus -1fil

\begin{IEEEbiography}[{\includegraphics[width=1in,height=1.25in,clip,keepaspectratio]{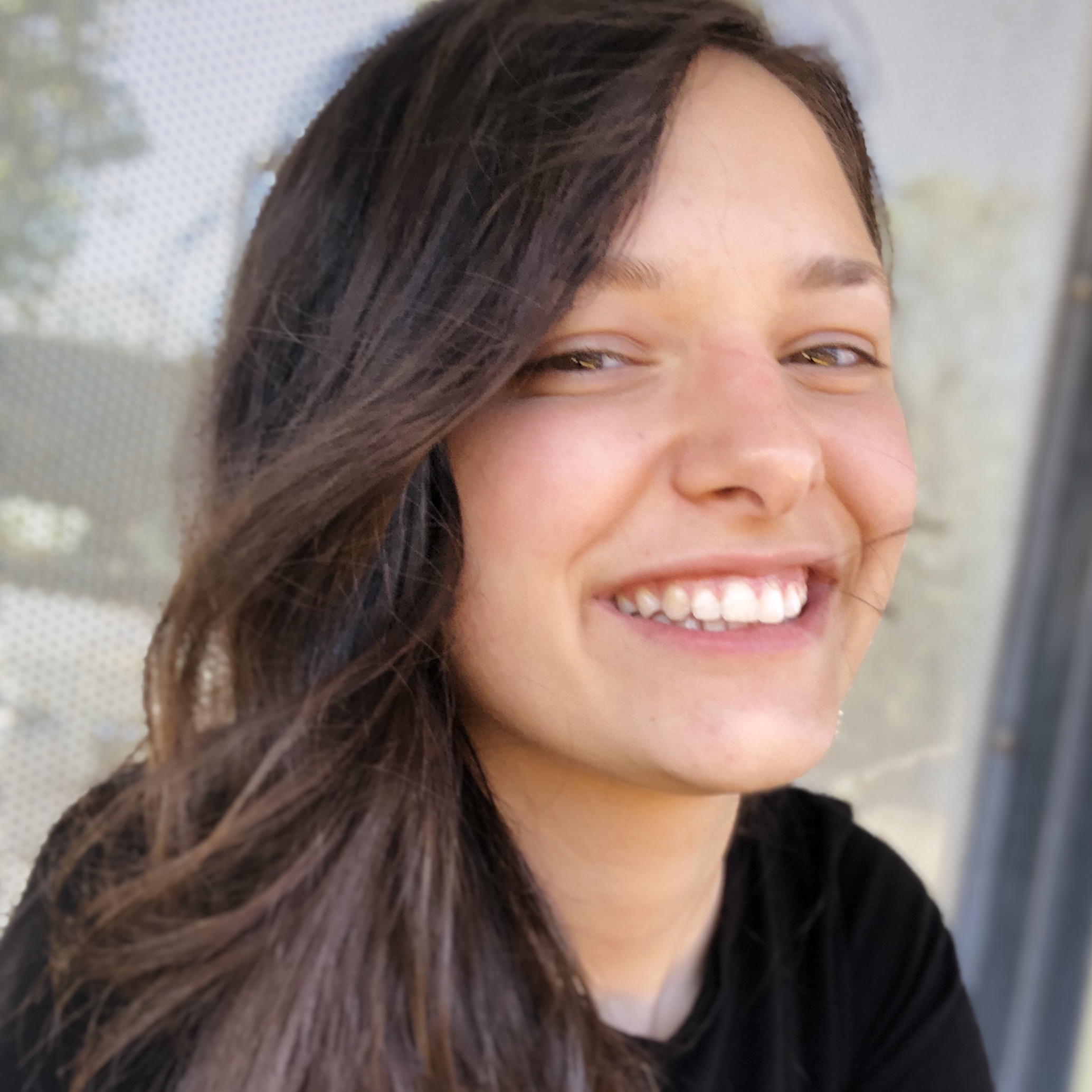}}]{Camelia D. Brumar}
is a Ph.D. candidate in Computer Science at Tufts University, where she also earned her M.S. Her research focuses on designing abstractions that help structure and reason about decision processes in visualization. Her research interests include problem definition in visualization workflows, frameworks for decision-making, and how these abstractions shape the design of visualization decision-support systems.
\end{IEEEbiography}



\enlargethispage{2\baselineskip}


\begin{IEEEbiography}[{\includegraphics[width=1in,height=1.25in,clip,keepaspectratio]{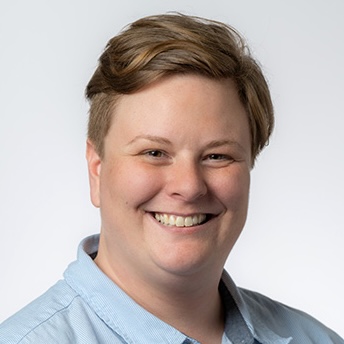}}]{Sam Molnar}
received their PhD in Computer Science at University of Colorado Boulder. They are a research scientist at the National Renewable Energy Laboratory. Their research work focuses on developing novel visualization designs and interaction techniques to support power system stakeholders’ analyses for many scenarios such as real-time operations, model comparisons, state estimation, science communication, etc.
\end{IEEEbiography}

\vfill



\begin{IEEEbiography}[{\includegraphics[width=1in,height=1.25in,clip,keepaspectratio]{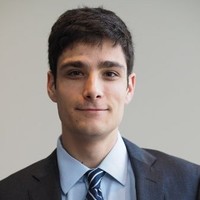}}]{Gabriel Appleby} 
received his PhD in Computer Science at Tufts University. He is a research scientist at the National Renewable Energy Laboratory. His research spans the fields of data visualization, visual analytics, and machine learning.
\end{IEEEbiography}
\vskip -2.5\baselineskip plus -1fil


\begin{IEEEbiography}[{\includegraphics[width=1in,height=1.25in,clip,keepaspectratio]{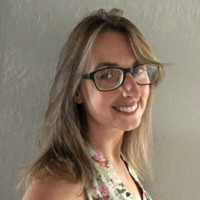}}]{Kristi Potter}
received her PhD in Computer Science at the University of Utah. She is a group manager and scientist at the National
Renewable Energy Laboratory. Her research is focused on methods to improve visualization techniques by adding qualitative information
regarding reliability to the data display. This work includes researching statistical measures of uncertainty, error, and confidence levels, and translating the semantic meaning of these measures into visual metaphors.
\end{IEEEbiography}
\vskip -2.5\baselineskip plus -1fil


\begin{IEEEbiography}[{\includegraphics[width=1in,height=1.25in,clip,keepaspectratio]{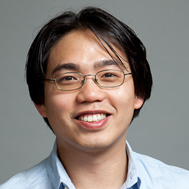}}]{Remco Chang} 
received his PhD in computer science from the University of North Carolina Charlotte. He is an associate professor in computer science with Tufts University. His research interests include
visual analytics, information visualization, human computer interaction, and databases.
\end{IEEEbiography}
\vskip -2.5\baselineskip plus -1fil




\newpage\hbox{}\thispagestyle{empty}\newpage


\end{document}